\documentclass[reprint,
superscriptaddress,gensymb,
amsmath,amssymb,aps,showkeys,showpacs,
twoside,final,secnumarabic,%raggedbottom,
nofootinbib]{revtex4-2}

\usepackage[paperwidth=205mm,paperheight=290mm,top=17mm,bottom=25mm,
inner=17mm,outer=17mm,
twoside]{geometry}

\usepackage{cmap} % Улучшенный поиск %русских слов в~полученном pdf-файле
\usepackage[T1,T2A]{fontenc}
\usepackage[utf8]{inputenc}
\usepackage[russian,english]{babel}
\usepackage{color}
\usepackage{graphicx}% Include figure files
\usepackage{dcolumn}% Align table columns on decimal point
\usepackage{bm} % bold math
\usepackage[unicode=true,colorlinks=true,linkcolor=magenta, urlcolor=blue, citecolor = blue,breaklinks]{hyperref}
\usepackage{multirow}
\usepackage{url}
\usepackage{breakurl}
\newcount\issue
\newcount\Vol
\newcount\numb
\usepackage{fancyhdr} %this packages %provides fancy up and bottom of page
\def\Vol{\textbf{78}}
\def\numb{x}
\begin{document}

%====== Начало шапки статьи  ============
\title{SPHERE-3: tackling the problem of primary cosmic ray\\mass composition with a new approach} 

\def\addressa{Faculty of Physics, M.V. Lomonosov Moscow State University\\ Leninskie gory, 1(3), Moscow, 119991, Russia.}
\def\addressb{Skobeltsyn Institute for Nuclear Physics, M.V. Lomonosov Moscow State University\\Leninskie gory, 1(3), Moscow, 119991, Russia.}
\def\addressc{Faculty of Computational Mathematics and Cybernetics, M.V. Lomonosov Moscow State University\\ Leninskie gory, 1(52), Moscow, 119991, Russia.}

\author{\firstname{V.I.}~\surname{Galkin}}
\email[E-mail: ]{v_i_galkin@mail.ru}
\affiliation{\addressa}
\affiliation{\addressb}
\author{\firstname{C.G.}~\surname{Azra}}
\affiliation{\addressa}
\affiliation{\addressb}
\author{\firstname{E.A.}~\surname{Bonvech}}
\affiliation{\addressa}
\author{\firstname{D.V.}~\surname{Chernov}}
\affiliation{\addressa}
\author{\firstname{E.L.}~\surname{Entina}}
\affiliation{\addressa}
\author{\firstname{V.I.}~\surname{Ivanov}}
\affiliation{\addressa}
\affiliation{\addressb}
\author{\firstname{V.S.}~\surname{Latypova}}
\affiliation{\addressa}
\author{\firstname{D.A.}~\surname{Podgrudkov}}
\affiliation{\addressa}
\affiliation{\addressb}
\author{\firstname{T.M.}~\surname{Roganova}}
\affiliation{\addressa}
\author{\firstname{M.A.}~\surname{Ziva}}
\affiliation{\addressc}

%\received{xx.xx.2023}
%\revised{xx.xx.2023}
%\accepted{xx.xx.2023}

\begin{abstract}
A new Cherenkov telescope of the SPHERE type is under development. Its main goal is to promote the solution of the problem of the primary cosmic ray mass composition at ultra high energies (1--100 PeV) using a newly developed technique of the primary mass assignment to EAS event on event-by-event basis. The telescope will carry out measurements of both the Cherenkov light reflected from the snow surface as well as the direct one. Sensitivity of the direct Cherenkov images' shapes to the primary mass is demonstrated. 
\end{abstract}

\pacs{29.40.Ka; 96.50.S-}\par
\keywords{primary cosmic rays, extensive air shower, Cherenkov light, dual detection  \\[5pt]}
%DOI:  

\maketitle
\thispagestyle{fancy}

%====== Начало  статьи  ============

\section{Introduction}\label{intro}

The long-dreamed 3D-detection of extensive air shower (EAS) may come true with a new SPHERE telescope designed for simultaneous direct and reflected Cherenkov light registration which will reduce substantially the uncertainties of the primary parameter estimates.

3D-detection of the super high energy cosmic ray in dense media became possible decades ago. We have a number of successful examples of detector arrays working together at different ground or water/ice levels: EAS-TOP \& MACRO at Gran Sasso Laboratory~\cite{infn} of INFN, DUMAND~\cite{DUMAND}, Baikal~\cite{Baikal}, AMANDA~\cite{AMANDA}, NESTOR~\cite{NESTOR}, ANTARES~\cite{ANTARES}, IceTop \& IceCube~\cite{IceCube}.
But we do not know any detector working at different levels in the atmosphere.

Generally speaking, obtaining the 3D information on EAS became available many years ago when it was first realized that the delay of Cherenkov photons can help to distinguish between the light coming from different stages of shower development. Still the relation between the emission altitude and the delay is approximate and sometimes even ambiguous, thus the real 3D-detection is preferable.
%(highly welcome).

SPHERE-3 telescope is going to establish a new era of 3D Cherenkov EAS detection with two synchronized telescopes, each pursuing its specific goal. A traditional mirror+PMT mosaic telescope will look for Cherenkov light reflected from the snowed surface, while a small lens+CCD one will register direct light angular image. Primary energy, direction and mass estimates will come from the well-known procedures dealing with the reflected CL image. If a direct CL image appears in coincidence with the reflected one, its processing provides an independent estimate of the direction. The shape of the direct image definitely contains some information on the mass of the primary particle.

SPHERE telescopes were conceived to contribute to the study of the primary cosmic ray (PCR) nuclei in the energy range $10^{15}-10^{18}$~eV which is only possible through the detection of the so called extensive air showers (EAS), i.e. huge particle cascades initiated by the primaries in the atmosphere. Of all primary parameters of EAS (primary energy, direction and particle type) the latter one turned out to be the most difficult to measure. Still its very important from the astrophysical viewpoint.
The super high energy PCR mass composition problem belongs to the particle astrophysics and first appeared at full strength after the physicists had become aware of a knee-like peculiarity in the PCR all-particle energy spectrum at about 3--5~PeV~\cite{knee}. Possible change of the mass composition presented one of the probable explanations for the fact. As of now many other features of the PCR energy spectrum are known and are waiting for the explanation.

\begin{figure*}[t]
\centering
\includegraphics[scale=0.8]{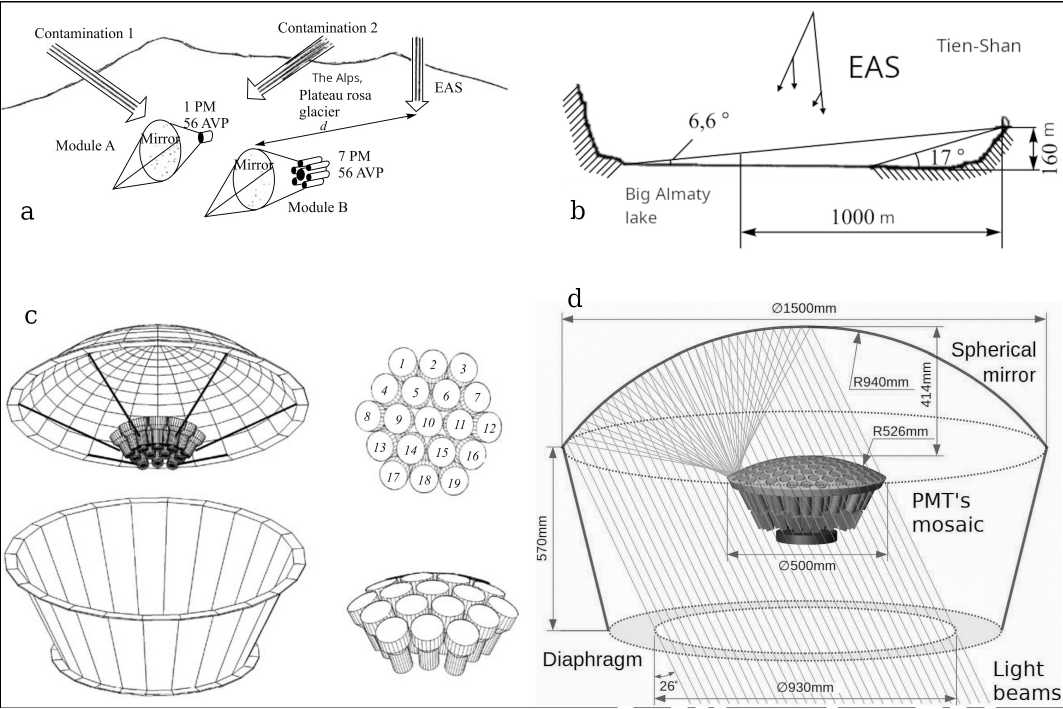}
\caption{Implementations of Chudakov's idea: a) in the Alps~\cite{Castagnoli1981}, b) in the Tien Shan~\cite{proto}, both ground-based; c) in the Volga region~\cite{SPHERE-1}, d) at lake Baikal~\cite{SPHERE-2}, both using tethered balloon as a carrier.}
\label{f1}
\end{figure*}

The original idea of the SPHERE method belongs to Alexander Evgenievich Chudakov~\cite{Chudakov1972}, a prominent Soviet cosmic ray researcher known also for many other important ideas and developments in the field. Chudakov suggested to collect the EAS Cherenkov light (CL) reflected by snowed ground with the use of an elevated telescope incorporating a mirror and a mosaic of PMTs. Such a device put a few hundred meters above the snow can observe a circle of ground of a similar radius and thus perform the function of a ground-based detector array of about the same size. The scheme presents some advantages as compared to the traditional arrays, the scalability of the experiment to be mentioned first.

The Chudakov's idea was implemented a few times (see Fig.1), first in 1980s as two ground-based telescopes observing the Plateau Rosa glacier in the Alps~\cite{Castagnoli1981} then in 1990s as a single telescope observing the snowed surface of the Big Almaty lake~\cite{proto}. The following two implementations, SPHERE-1~\cite{SPHERE-1} and SPHERE-2~\cite{SPHERE-2} used tethered balloons as carriers and included rather detailed PMT mosaics. The latter 
one was the most advanced and succeeded in yielding some important results on the PCR energy spectrum and mass composition at about 10 PeV ~\cite{res}. The processing technique was a rather complicated one but the full simulation of the experiment has not been carried out before the measurements. Only now becomes it possible to simulate the registration of EAS CL by the SPHERE-2 telescope in proper detail and we have learned much by  its results~\cite{detcalc}. Particularly, we realized that the sensitivity of SPHERE method to the primary mass, which is one of its main advantages, should be the main idea of the SPHERE design.
We have developed an approach to the processing of a shape of the reflected CL image on SPHERE-2 mosaic that makes it possible to estimate the mean primary mass dependence on the primary energy~\cite{Vas}. But the error of the mass estimates for individual EAS is still rather large and this should be taken into account while designing a new SPHERE-3 experiment.

The data of SPHERE-2 experiment gave us a hint at how to gather more information of a shower. Some experimental data frames contain signals unusual for the reflected EAS CL which are too short and precede the target signal by a few microseconds. Dedicated studies identify the features as the signals of the direct CL photons reaching the mosaic PMTs through some holes in the mirror. Fig.2 shows an example of such a data frame: the delay between the unusual signal and the usual one (about 4 $\mu$s) corresponds to the approximately doubled flight altitude of the SPHERE-2.

\begin{figure}[t]
\centering
\includegraphics[scale=0.7]{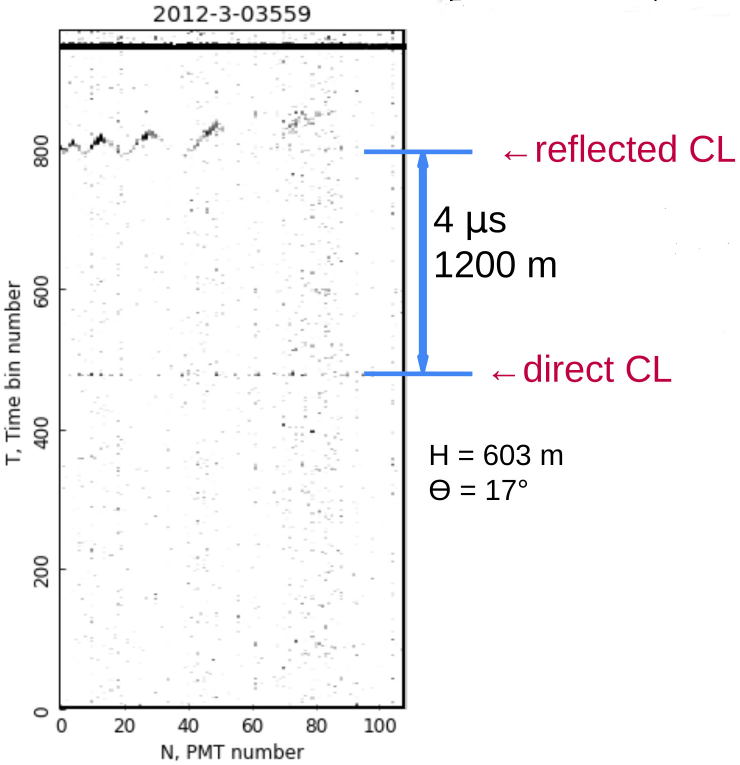}
\caption{During the SPHERE-2 data analysis an unusual event was discovered with the usual reflected Cherenkov image preceded by an unusual one identified as a direct image.}
\label{f2}
\end{figure}

\section{The goal of the new project: optimization of the detector design to reduce the primary mass estimate uncertainty}

Any EAS detector array has to estimate the primary parameters of a shower (primary particle energy, direction and mass) as best as possible. Chudakov's flying telescope (let us call it SPHERE hereafter) is expected to do that as well. It even has some advantages over the ground-based detectors.

First, the very idea of such experiment suggests its scalability: one and the same telescope can probe different primary energy ranges depending on the observation altitude. With tethered balloon as a carrier the telescope can be elevated to 1 km and even higher so that it can observe an area of about 1 km${}^2$ which rivals the areas of the ground-based EAS detector array studying 1--1000 PeV energy range. The higher it flies the higher are the primary energies considered.

Second, SPHERE telescopes are capable of observing the major part of the ground area within the field of view unlike the ground-based installations with detectors put far apart (tens to hundreds m). Here one can anticipate an invaluable new property: the vicinity of the shower axis can be seen and analyzed, which is an important region from the viewpoint of the primary mass estimation.

As of now the SPHERE telescope history exceeds 40 years, still the results of its work could be more impressive with the advantages mentioned. In our viewpoint, there are two main obstacles to overcome while studying the PCR mass composition.

The first one is the general problem of cascade fluctuations which smear the beautiful mean characteristics of the cascade showers commonly used to deduce the primary nucleus mass. While the primary energy and direction can be estimated with the relatively straightforward methods, the primary mass reveals itself in the details of the differential distributions of observable characteristics which are substantially affected by the fluctuations. These details require special attention as they differ for different detectors and should be measured and processed carefully. The latter circumstance may be called an information greed of the primary mass, which is usually ignored when planning the experiment. More accurately, all the primary parameters are thought to be about equally greedy for information and no attempts are made to favor the primary mass. Thus, the way to overcome the first obstacle is to admit the information greed of the primary mass and take it into account while constructing the detector.

The second obstacle to overcome is the absence of general approach to find the primary mass signature for the concrete type of experiment on condition that one doesn't know the details of the hadron interaction at super high energies. We think we have already found such an approach which can lead to mass-sensitive parameters with minimum dependence on the interaction model: one must use the distribution shape parameters for the primary mass estimation~\cite{Vas}.

We are going to bypass both obstacles in a new SPHERE-3 project. Hopefully, we will work out some new principles of such a bypass which could be useful for the other EAS Cherenkov detector arrays.

Thus, we can state the following main goal of the new project: to optimize the telescope design so that to ensure minimum uncertainty of the primary mass estimates.

\section{Where do we stand in the primary mass estimation problem?}

The data of the previous SPHERE-2 experiment are being thoroughly reanalyzed at the moment using a new vast database of artificial EAS events. The new analysis allows to draw some conclusions on the properties of the past telescope that must be improved in the future one.

We constructed quite a number of mass-sensitive parameters describing the shape of reflected CL images in SPHERE-2 telescope, selected the best ones (those showing minimum classification errors for two pairs of primary nuclei, see~Fig.3) and assessed the primary mass estimation errors in individual EAS and on average for different mass compositions with sample volumes of a few thousand events. Individual errors are not at all satisfactory (they can amount to hundreds percent) but the mean ones are good enough: $\sim2 \%$.

\begin{figure}[t]
\centering
\includegraphics[scale=0.6]{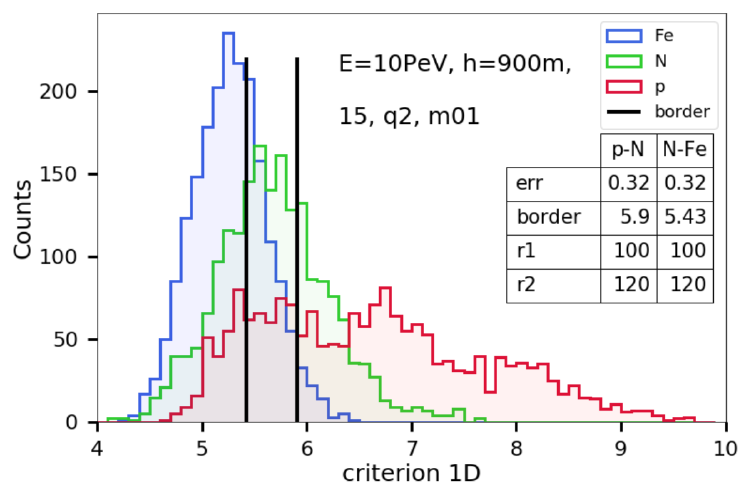}
\caption{Primary mass classification results using the reflected CL images in SPHERE-2. Classification errors (designated as {\bf err}) were used to choose the optimum criterion parameter. They are rather high. Shown are 10 PeV EAS with $\theta = 15^\circ$ seen from 900~m altitude. Interaction model QGSJETII-04. Standard atmosphere by Linsley.}
\label{f3}
\end{figure}

Thus, one has to acknowledge that the shape of the reflected CL image is sensitive to the primary mass and it is worth while to extract the appropriate information in the best possible way. To do so we are going to check the capabilities of a few candidate designs of the telescope in this respect. At this point one should recall the possibility to carry out a dual detection of a shower using both the reflected and the direct CL. Of course, this can't happen with the balloon-borne telescope but we are going to use an unmanned aerial vehicle as a carrier in the future. Fig.4 shows the detection scheme for this case.

\begin{figure}[t]
\centering
\includegraphics[scale=0.6]{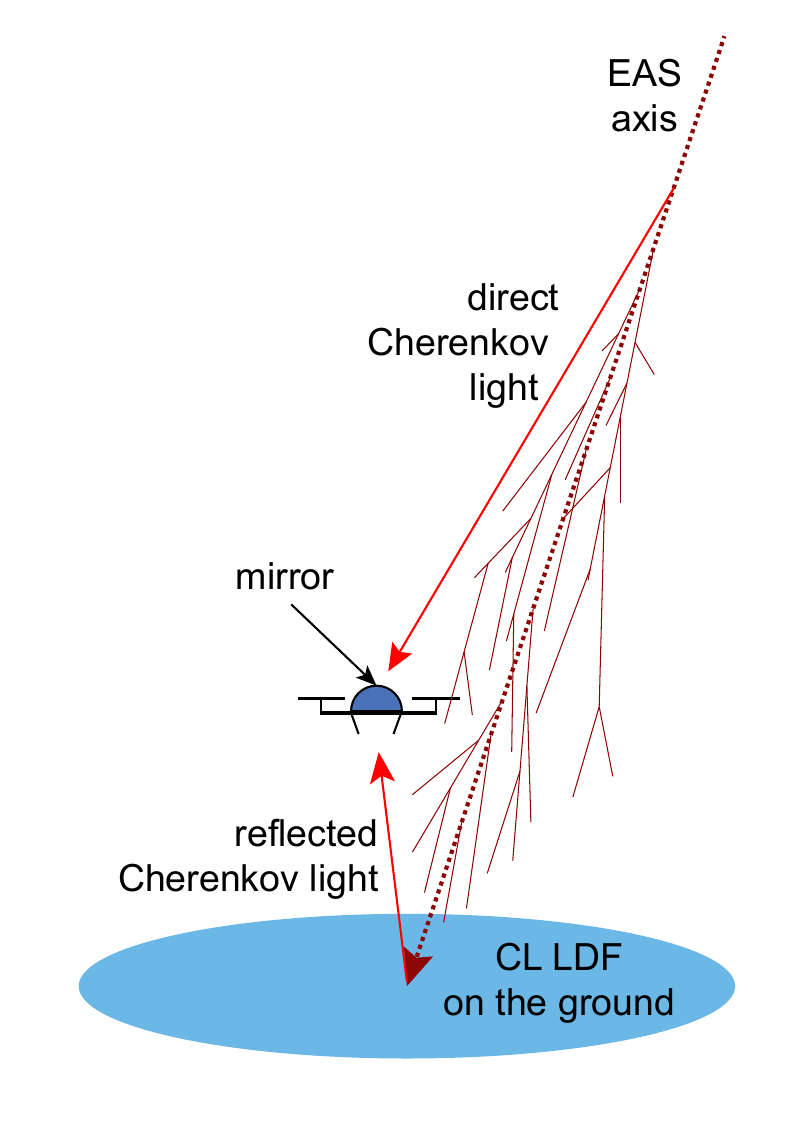}
\caption{Dual detection scheme using reflected and direct EAS CL.}
\label{f4}
\end{figure}

It is common knowledge that the angular distribution of EAS direct CL contains important information on the primary mass. It is enough to mention the success of the Cherenkov $\gamma$-ray astronomy is mostly due to this fact~\cite{Hillas1985}. Still we can add our own developments pertinent to the PCR mass composition problem~\cite{Pamir-XXI}. It turned out that the super high energy showers from the primary nuclei can yield much richer information that the usual $\gamma$-ray showers of TeV energy range. Thus, one can use much more advanced CL angular distribution shape parameters instead of the usual Hillas set~\cite{NIMA}. With the direct CL it is also possible to find the shape parameters sensitive to the primary mass but almost independent of the hadron interaction model.

We are going to combine the capabilities of the reflected and the direct CL where possible in order to reduce the uncertainties of the primary mass estimates. It is important to understand the optimum detection altitudes for both reflected and direct CL. For the purpose vast simulations are underway of artificial EAS events with CORSIKA~\cite{CORSIKA} code resulting in EAS CL characteristics at four levels: lake Baikal level, 0.5, 1.0 and 2.0 km above. Next simulation stage will use this database to produce reflected and direct CL images at three flying altitudes (0.5, 1.0 and 2.0 km). Generally speaking, we have already learned much about the reflected image processing while looking for the mass-sensitive parameters for SPHERE-2 but we are not decided yet on the direct CL detector and therefore we present here only the simplest evidence of the direct CL sensitivity to the primary mass.

\begin{figure}[t]
\centering
\includegraphics[scale=0.72]{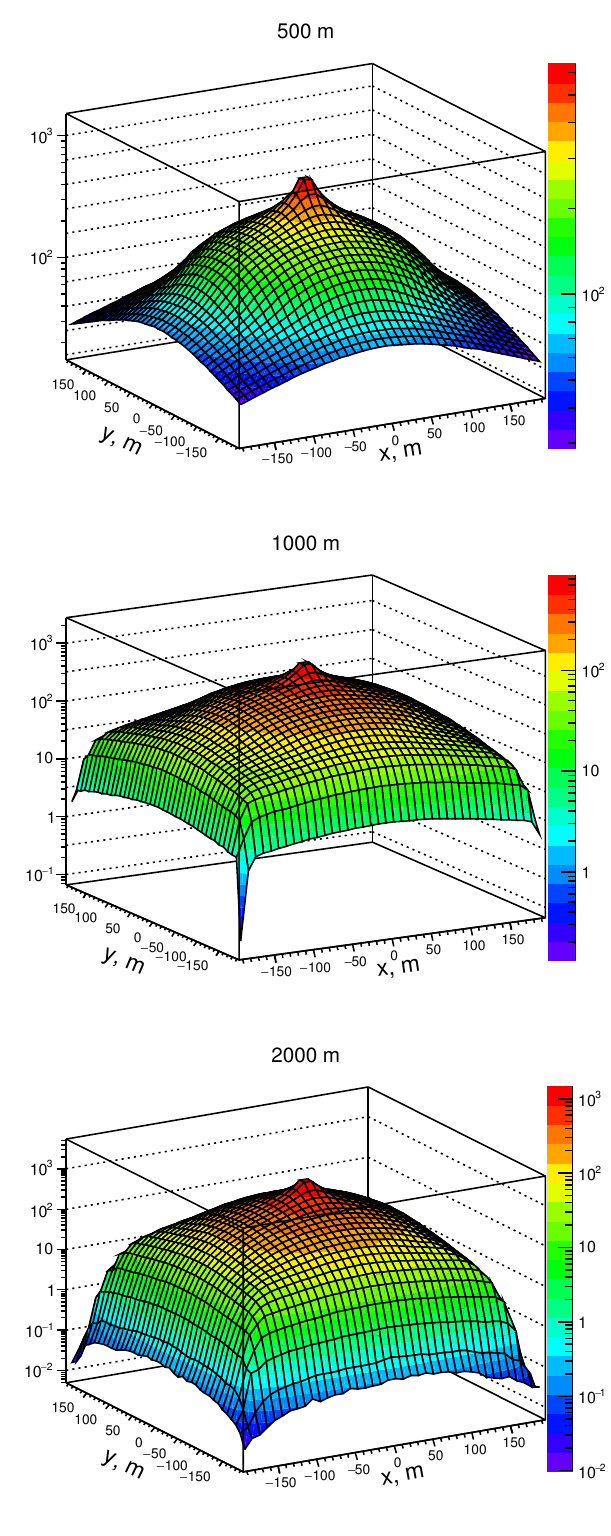}
\caption{Lateral distribution of photoelectrons born by Cherenkov photons of an EAS caused by 1 PeV proton. Vertical axes: photoelectron per detector area. Detector parameters: see text.}
\label{f5}
\end{figure}

\begin{figure}[t]
\centering
\includegraphics[scale=0.72]{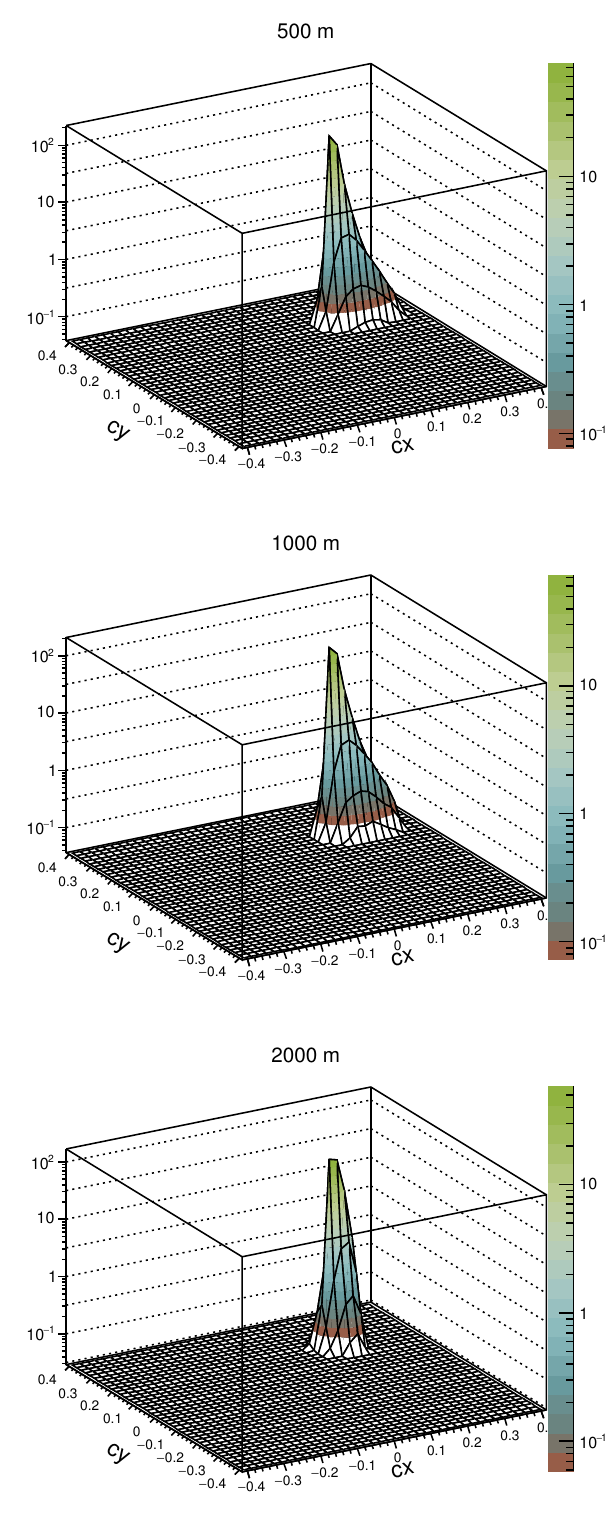}
\caption{Angular distribution of photoelectrons born by Cherenkov photons of an EAS caused by 1 PeV proton at 100 m core distance. Vertical axes: photoelectron per detector area per pixel (about $1^\circ \times 1^\circ$). Detector parameters: see text.}
\label{f6}
\end{figure}

\begin{figure}[t]
\centering
\includegraphics[scale=0.72]{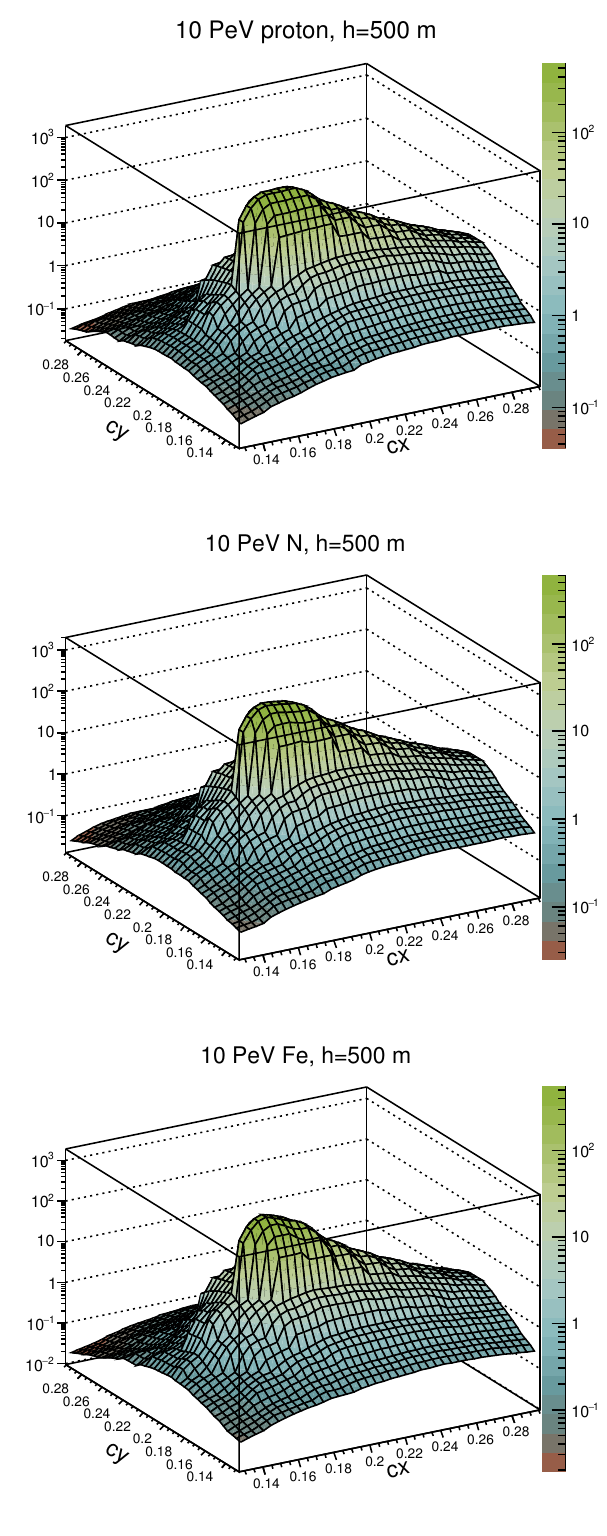}
\caption{Angular distributions of photoelectrons born by Cherenkov photons of an EAS caused by 10 PeV proton, nitrogen and iron nuclei at 100 m core distance. Observation altitude 500 m. Vertical axes: photoelectron per detector area per pixel (about $1^\circ \times 1^\circ$).}
\label{f7}
\end{figure}

%\vspace*{0.4cm}
%{separation_10PeV_500m}
\begin{figure}[!t]
\centering
\includegraphics[scale=0.45]
{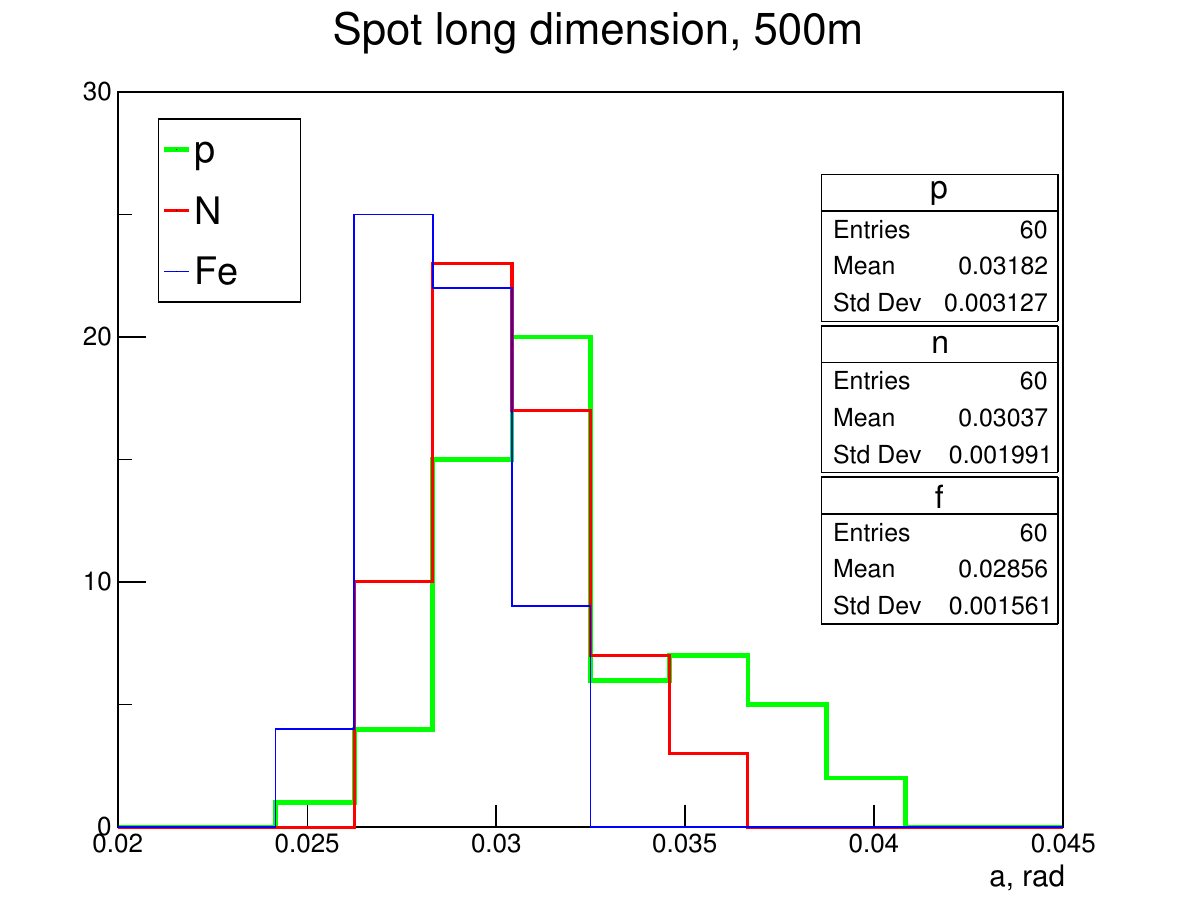}
\caption{Long axis length {\bf a} distributions of the direct CL images of EAS caused by 10 PeV proton, nitrogen and iron nuclei at 100 m core distance. Observation altitude 500 m.}
\label{f8}
\end{figure}

\section{Direct CL can help to estimate the primary mass}

Here we assume the area of the direct CL detector (probably, CCD sensor + lens) to be 1 dm${}^2$, its field of view will be $50^\circ \times 50^\circ$ with the axis aimed at zenith.

First, one must understand the size of the shower axis vicinity where such a detector can see the shower image. Lateral distribution of CL with photons within the field of view solves the problem (Fig.5). One can conclude that even a proton shower of energy 1 PeV can fill the detector with $\sim$100 photoelectrons at core distances up to 150 m at 500 m altitude, up to 140 m at 1000 m altitude and up to 110 m at 2000 m altitude.

Second, one must see the angular distribution and choose the shape parameter to be used for the mass estimation. Fig.6 presents CL angular distributions for the three altitudes. One can see sharp peaks pointing approximately in the shower arrival direction ($\theta = 15^\circ$), the higher is the observation level the sharper is the peak. It is not at all easy to decide on the mass sensitive parameter so we try the first one we can think of which is one of the Hillas set: the angular peak long dimension {\bf a}. Later a thorough analysis is to be made to choose the most sensitive parameter. Now we just demonstrate that such an analysis is worth while.

Obviously, one should not use the shape of the whole distribution because its periphery will be affected by the fluctuations and the background. We now set the threshold to 1\% of the maximum pixel content, to be tuned later on. Fig.7 shows the tops of the direct CL angular distributions for three 10 PeV nuclei (proton, $^{14}$N, $^{56}$Fe), core distance 100 m, observation altitude 500 m. Their look very similar, still their long axes differ in size and this can be used to distinguish between the nuclei. Angular distributions for 1000 m altitude behave similarly while at 2000 m the scale of the long axes is much smaller which makes the images of different nuclei more difficult to distinguish, higher angular resolution is required. Fig.8 presents the histograms of the long axis sizes of the angular distributions shown in Fig.7. It is clear that the distributions are really different but the classification errors for pairs p-N and N-Fe are rather high (Table 1). It should be noted that the mass sensitive parameter $a$ was chosen almost at random and still demonstrates a sensitivity to the primary mass comparable to that obtained with the optimized parameter selected for SPHERE-2 data (Fig.3).

Certainly, a more sensitive parameter will be found while analyzing the bulk of the SPHERE-3 artificial image database.
One may learn an interesting tendency from Table1: higher observation levels favor the separation of massive nuclei while the light ones are separated better at 500 or 1000 m. It is because the differences of the cascade curves for the heavier primaries appear higher than for the light ones.

\begin{table}[h]
\begin{center}\caption{Missclassification probabilities using parameter {\bf a}. Sample volume 60 for each nucleus.}
%\begin{ruledtabular}
\begin{tabular}{|c|c|c|c|}\hline
%\begin{tabular}{|p{0.8in}|p{1.0in}|p{1.0in}|} \hline
$E_0$, & $H_{obs},$ & \multicolumn{2}{c |}{Pair of nuclei} \\ \cline{3-4}
{ PeV}    & {m}     & p-N & N-Fe \\ \hline
10 & 500 & 0.35/0.40 & 0.30/0.35 \\ \hline
10 & 1000 & 0.35/0.32 & 0.37/0.27 \\ \hline
10 & 2000 & 0.35/0.40 & 0.30/0.35 \\ \hline
30 & 500 & 0.33/0.35 & 0.30/0.27 \\ \hline
30 & 1000 & 0.33/0.30 & 0.32/0.25 \\ \hline
30 & 2000 & 0.35/0.40 & 0.35/0.23 \\ \hline
\end{tabular}\label{tab:table1}
\end{center}
\end{table}

\section{Conclusions}

The main goal of the current stage of our work is to find the optimal design of
the new telescope with respect to the primary mass resolution.

We have already developed a procedure for such optimization of the traditional part of the SPHERE telescope acquiring the reflected CL.
Now that we have perceived the possibility to detect also the direct CL, i.e. to
ensure the EAS detection at two levels, we are going to extend the optimization to the upper part of the telescope.

Direct CL of EAS definitely shows sensitivity toward the primary mass.
Hopefully, combined with the sensitivity provided by the reflected CL (e.g., the reflected
CL image steepness parameter) it can substantially reduce the uncertainty of
the individual shower primary mass estimate.

In case we succeed in developing this promising technique one may call it the first realization of the EAS  three-dimensional detection.

\section*{ACKNOWLEDGMENTS}
This work was supported by a grant from the Russian Science Foundation No 23-72-00006, https://rscf.ru/project/23-72-00006. 

The research is carried out using the equipment of the shared research facilities of HPC computing resources at Lomonosov Moscow State University~\cite{Lom-2}.

\section*{CONFLICT OF INTEREST}
The authors declare that they have no conflicts of interest.

\end{document}